\documentclass[11pt,twoside]{./atmp}
\usepackage{amssymb}

\usepackage{amsmath,amssymb}
\usepackage{color}

\copyrightnotice{2007}{xx}{1}{20} 
\input{tcilatex}

\arxurl{gr-qc/0701174}

\begin{document}

\title[One common solution ...]
{One common solution to the singularity and perihelion problems}

\author{Branko Sari\'{c}}

\address{32 000 \v{C}a\v{c}ak, Kralja Petra I br.1, Serbia}\ 
\email{bsaric@ptt.yu}

\begin{abstract}
Based on the kinetic energy theorem, as one of the fundamental theorems from
the classical mechanics, throughout the first part of the article an attempt
has been made to derive the mathematical model of a material point motion in
the three-dimensional spatial subspace of the integral four-dimensional
space-time continuum and in the field of action of an active force $\mathbf{F%
}$. Accordingly, with a view to surmounting the singularity problem on the
one hand, as well as the moving perihelion problem of the planets on the
other, as two acutely vexed questions within \textit{Newton's} gravity
concept, the paper ends with modification of \textit{Newton's} gravity
concept itself.
\end{abstract}

\maketitle

\cutpage

\setcounter{page}{2}

\noindent

\section{Introduction}

The space-time mathematical model of motion of a material point in the
relativistic (\textit{Einstein's}) mechanics, based on the principle of the
constancy of the light velocity in vacuum relative to inertial frames \cite
{Gr,Lu,Pa}, is said to be approximately more general with respect to the so
called spatial mathematical model derived in the classical (or \textit{%
Newton's}) mechanics \cite{An,Mi}. The four-dimensional space-time continuum
a priori\ established by \textit{Minkowski} is the foundation stone of whole
relativistic mechanics. Hence, with a view to pointing out the physical
sense of a configurative space of the space-time continuum, throughout the
first part of the paper an attempt has been made to derive, on the basis of
some fundamental principles of the mechanics, the mathematical model of a
material point motion in three-dimensional spatial subspace of an ambient
integral four-dimensional space-time continuum.

On the other hand, \textit{Newton's} gravity concept, in existence for
already three centuries, which describes with sufficiently exactness, in
spite of some acutely vexed questions within it, the so called \textit{Sun's}
planetary system via \textit{Kepler's} laws of a planetary motion, is one of
the fundamental laws of the classical mechanics, particularly of the
celestial mechanics. The first vexed question, based on the purely
theoretical basis, is the so-called singularity problem. Namely, on the
basis of the mathematical model of two material points motion of the same
mass in the field of action of the central \textit{Newton's} gravity force,
when the direction of material points motion coincides with the assaulted
direction of the force \cite{Mi,Ra}, it is easy to see that absolute values
of all relevant physical variables, such as velocity, force, kinetic and
potential energy, in the limit as mutual distance of the material points
tends to zero, tend to infinity. The second one, which is cleanly empirical
nature, is the perihelion problem. Namely, it has been experimentally stated
that the perihelion of \textit{Mercury's} orbits moves into the plane of its
planetary motion around the \textit{Sun}. In other words, all planetary
motions of \textit{Sun's} planetary system depart from elliptical orbits
obtained from \textit{Newton's} mathematical gravity model \cite{Lu,Ta}. By
the strict \textit{Schwarzshild-Droste's} solution to the static
gravitational field with spherical symmetry, in the general \textit{%
Einstein's} relativity theory \cite{Pa}, the perihelion problem has been
approximately solved. However, this solution does not solve the singularity
problem. Accordingly, to solve simultaneously these two acutely vexed
questions within \textit{Newton's} gravity concept the manuscript ends with
an\ approximative modification of \textit{Newton's} gravity concept itself.

\subsection{Fundamental characteristics of the space-time continuum}

By the notion of a material point, introduced for the purpose of an useful
idealization, as one of the underlying notions of physics in the general
sense, but not only of physics, one means a geometrical point, which is
spatially no dimensional on the one hand and exactly fixed mass on the other 
\cite{Mi}. Closely related to the geometrical point notion is the set of
values$\,\left( a_{1},a_{2},...,a_{\alpha }\right) $ of some arbitrary $n$
variables $\left( x_{1},x_{2},...,x_{\alpha }\right) $, such that a set of
all geometrical points, and for all real values of the variables, is the
real $n$ - dimensional configurative space \cite{An}. The geometrical point
defined by a set of zero values $\left( 0,0,...,0\right) $ is the zero
co-ordinate point. If and only if one of $n$ arbitrary variables $x_{\alpha
} $ ($\,\alpha =1,2,...,n$) is the time variable $t$, the space
aforementioned becomes the space-time continuum (the integral space) \cite
{Gr,Iv,Ta}. The value of $t$ is called moment or instant \cite{Gr}.

The set of all geometrical points of the spatial subspace of the integral
space, to which the mass $m$ can be joined in some strictly monotonous
sequence of permitted instants of the time $t$, makes an odograph, that is a
trajectory (a path of motion) of the material point $\mathcal{M}$. The time
variable $t$ is taken for a unique independent variable \cite{Gr}, so that
all remaining spatial variables $x^{i}\left( t\right) $ are functional
variables. The set of all geometrical points $x^{\alpha }\left( t\right) $%
\footnote{{\footnotesize \textit{Greek} indices take values }$1,2,...,n$%
{\footnotesize , and \textit{Latin} ones }$1,2,...,n-1${\footnotesize .}} of
the integral space is an integral curve of $\mathcal{M} $ and an odograph,
that is a trajectory (a path of motion) of a representative point $\mathcal{%
\hat{M}}$ ($m=\hat{m}$) of the space-time continuum.

The vectors $\mathbf{r}\left[ x^{i}\left( t\right) \right] $ and $\rho \left[
x^{\alpha }\left( t\right) \right] $ defined with respect to the origin are
position vectors of $\mathcal{M}$ and $\mathcal{\hat{M}}$, respectively, in
the configurative space of the space-time continuum. The concept of a vector
in vector hyper-dimensional spaces ($n>3$) should be conditionally
comprehended in the sense of its geometrical presentation in a form of
segments. Hence it bears a name linear tensor \cite{Pa}. The set of
infinitesimal values $dx^{\alpha }$ ($\alpha =1,2,...,n$) corresponds to
infinitesimal movements of $\mathcal{M}$ along the trajectory of motion.

Covariant vectors $e_{\alpha }=\partial _{\alpha }\rho \left( x^{\beta
}\right) $, where $\partial _{\alpha }$ denotes $\partial /\partial
x^{\alpha }$, form a covariant vector basis $\left\{ e_{\alpha }\right\}
_{\alpha =1}^{n}$ of the configurative space. The vectors $e^{\beta }$, such
that at any point of the space $e_{\alpha }\cdot e^{\beta }=\delta _{\alpha
}^{\beta }$, where the second order system of the unit values $\delta
_{\alpha }^{\beta }$ is the identity $n\times n$ matrix (\textit{Kronecker's}
delta-symbol) \cite{An,L-L,Pa}, form a basis $\left\{ e^{\beta }\right\}
_{\beta =1}^{n}$ which is called a dual basis of the covariant vector basis $%
\left\{ e_{\alpha }\right\} _{\alpha =1}^{n}$. This is so-called natural
isomorphism from $\left\{ e_{\alpha }\right\} _{\alpha =1}^{n}$ onto $%
\left\{ e^{\beta }\right\} _{\beta =1}^{n}$. Accordingly, the differential\ $%
d\rho $ of the position vector $\rho $ of $\hat{m}$ is defined by $d\rho
=dx^{\alpha }e_{\alpha }=dx_{\beta }e^{\beta }$, where the so called \textit{%
Einstein's} convention is applied to a summation with respect to the
repetitive indexes (uppers and lowers) \cite{An,Pa}, herein as well as in
the further text of the paper.

\section{Main results}

\subsection{A metric of configurative spaces}

On the basis of preceding description of the space-time continuum one can
conclude that the space-time continuum is a metrical affine vector space,
whose linearly independent basic (fundamental) co-ordinate vectors, reduced
to the origin, form a basic $n$ - hedral \cite{An}. As for a metric of
metrical space of the space-time continuum it is explicitly related to
fundamental properties of a material point motion. Namely, on the one hand
the basic kinematical characteristic of a material point motion is the
velocity $\mathbf{v}=d_{t}\mathbf{r}$, where $d_{t}$ denotes $d/dt$, while
on the other hand the basic dynamical characteristics are the quantity of
movement $\mathbf{K}=$ $m\mathbf{v}$ and the kinetic energy $\mathcal{E}%
=m\left( \mathbf{v}\cdot \mathbf{v}\right) /2=mv^{2}/2$ \cite{Mi,Sa 3}.

The functional expressions for kinetic energies $\mathcal{E}$ (of $\mathcal{%
M)}$ and $\mathcal{K}$ (of $\mathcal{\hat{M})}$ can be stated in more
appropriate forms: 
\begin{equation}
\frac{2}{m}\mathcal{E}\left( dtdt\right) =d\mathbf{r}\cdot d\mathbf{r}%
=\left( ds\right) ^{2}  \label{1}
\end{equation}
and 
\begin{equation}
\frac{2}{\hat{m}}\mathcal{K}\left( dtdt\right) =d\rho \cdot d\rho =\left(
d\sigma \right) ^{2}\text{,}  \label{2}
\end{equation}
where $ds$ and $d\sigma $ are line elements of affine metrical spaces of the
spatial and space-time continuum, respectively \cite{An}. On the other hand,
according to the differential equation of energy balance of $\mathcal{M}$ 
\cite{AN,Sa 1} (the differential equation of \textit{the kinetic energy
theorem} \cite{Mi}): 
\begin{equation}
d(\frac{1}{2}m\mathbf{v}\cdot \mathbf{v)}=d\mathcal{E}=\mathbf{F}\cdot d%
\mathbf{r}=d\mathcal{A}\text{,}  \label{3}
\end{equation}
derived from the second \textit{Newton's} low (principle) of a material
point motion in the field of action of an active force $\mathbf{F}$ \cite
{Mi,N-V}: 
\begin{equation}
md_{t}\mathbf{v}=\mathbf{F}\text{,}  \label{4}
\end{equation}
an infinitesimal change of $\mathcal{E}$ is equal to an infinitesimal work $d%
\mathcal{A}$ of $\mathbf{F}$ during infinitesimal movements of $\mathcal{M}$
along the path of motion, that is 
\begin{equation}
d\left( \mathcal{E}-\mathcal{A}\right) =0\text{.}  \label{5}
\end{equation}

If an additive integral constant $\hat{k}$ of (\ref{5}) is introduced into
the analysis: 
\begin{equation}
\mathcal{E}-\mathcal{A}=\hat{k}=\frac{1}{2}m\hat{c}^{2}\text{,}  \label{6}
\end{equation}
then, by (\ref{1}) and (\ref{6}), it follows that 
\begin{equation}
d_{t}\mathbf{r}\cdot d_{t}\mathbf{r}-\hat{c}^{2}=\frac{2}{m}\mathcal{A}\text{%
,}  \label{7}
\end{equation}
that is 
\begin{equation}
\left( ds\right) ^{2}-\hat{c}^{2}\left( dt\right) ^{2}=\frac{2}{m}\mathcal{A}%
\left( dt\right) ^{2}\text{.}  \label{8}
\end{equation}

\begin{description}
\item[Assumption 2.1]  \label{Pr}\textit{An integral curve of }$M$\textit{\
in the space-time continuum is a curve in the four-dimensional space-time
continuum of Minkowski}.$\blacktriangledown $
\end{description}

On the basis of the preceding assumption and the fact that 
\begin{equation}
\left( ds\right) ^{2}-c^{2}\left( dt\right) ^{2}=\left( d\sigma \right) ^{2}%
\text{,}  \label{9}
\end{equation}
in the space-time continuum of \textit{Minkowski} \cite{Pa,Lu}, where the
constant $c$ is nominally equal to the light velocity in vacuum, it follows
that 
\begin{equation}
\mathcal{K}=\frac{1}{2}\hat{m}\left( d_{t}\rho \cdot d_{t}\rho \right) =%
\mathcal{A}+\bar{k}\text{,}  \label{10}
\end{equation}
where $\bar{k}$ is an additive integral constant of $d\left( \mathcal{K}-%
\mathcal{A}\right) =0$, that is 
\begin{equation}
\mathcal{E}-\mathcal{A}=\hat{k}=\frac{1}{2}mc^{2}+\bar{k}  \label{11}
\end{equation}
taking the relation (\ref{6}) into account. In addition $\hat{c}^{2}=c^{2}+2%
\bar{k}/m$, since $m=\hat{m}$.

In the event that the \textit{Pfaff} form $\mathbf{F}\cdot d\mathbf{r}$ of (%
\ref{3}) is absolute differential, in other words if there exists a scalar
valued function $\mathcal{P}$ depending on $\mathbf{r}$ such that $\mathbf{F}%
=-grad\mathcal{P}\left( \mathbf{r}\right) $, that is $d\mathcal{A}=-d%
\mathcal{P}$, then a material point motion is that in the field of action of
an active potential force $\mathbf{F}$ with the potential $\mathcal{P}$ \cite
{An,Mi}. Accordingly, 
\begin{equation}
\mathcal{U}=\mathcal{E}+\mathcal{P}  \label{12}
\end{equation}
is a functional of the total mechanical energy of $\mathcal{M}$. Heaving in
view the fact that 
\begin{equation}
d\left( \mathcal{E}+\mathcal{P}\right) =0\text{,}  \label{13}
\end{equation}
$\mathcal{U}$ is an integral of a motion of $\mathcal{M}$ too \cite{Mi}.

In the general case of a material point motion in the field of action of an
active potential force $\mathbf{F}$, it follows immediately from the
differential form of the energy conservation, see (\ref{13}), that 
\begin{equation}
\mathcal{E}+\mathcal{P}=\mathcal{U}=\hat{k}+\tilde{k}=\frac{1}{2}mc^{2}+(%
\bar{k}+\tilde{k})=k+%
\kappa%
\text{,}  \label{14}
\end{equation}
where $\tilde{k}$ is an additive integral constant of $d\mathcal{A}=-d%
\mathcal{P}$ and $k=mc^{2}/2$ as well as $%
\kappa%
=\bar{k}+\tilde{k}$.

Note that the metric form $\left( d\sigma \right) ^{2}=d\rho \cdot d\rho $
of the configurative space has been adopted in such a way to represent 
\textit{the kinetic energy theorem} on the one hand and \textit{Assumption 
\ref{Pr}} on the other. Accordingly, one can say that in that a way it is
possible to reduce the analysis of a material point motion to the analysis
of a representative point motion in the configurative space. Namely, if one
starts with the action $\mathcal{S}$ in the \textit{Lagrange} sense along
the path of $\mathcal{\hat{M}}$ in the configurative space \cite{An}: 
\begin{equation}
\mathcal{S}=2\int_{t_{2}}^{t_{1}}\mathcal{K}dt\text{,}  \label{15}
\end{equation}
then it follows from the relations (\ref{1}) and (\ref{10}) that 
\begin{equation}
\mathcal{K}=\bar{k}+\tilde{k}-\mathcal{P}=%
\kappa%
-\mathcal{P}=\frac{1}{2}m\left( d_{t}\sigma \right) ^{2}\text{,}  \label{16}
\end{equation}
that is 
\begin{equation}
\mathcal{S}=\sqrt{2m}\int_{\sigma \left( t_{1}\right) }^{\sigma \left(
t_{2}\right) }\sqrt{\mathcal{K}}d\sigma =\sqrt{2m}\int_{\sigma \left(
t_{1}\right) }^{\sigma \left( t_{2}\right) }\sqrt{%
\kappa%
-\mathcal{P}}d\sigma \text{.}  \label{17}
\end{equation}

In that case an action line element $dw$ is introduced \cite{An} as follows 
\begin{equation}
\sqrt{\frac{1}{2}mc^{2}}dw=\sqrt{%
\kappa%
-\mathcal{P}}d\sigma \text{,}  \label{18}
\end{equation}
so that the action metric form of the configurative space is 
\begin{equation}
k\left( dw\right) ^{2}=ka_{\alpha \beta }dx^{\alpha }dx^{\beta }=\left( 
\kappa%
-\mathcal{P}\right) e_{\alpha \beta }dx^{\alpha }dx^{\beta }=\left( 
\kappa%
-\mathcal{P}\right) \left( d\sigma \right) ^{2}\text{,}  \label{19}
\end{equation}
where $e_{\alpha \beta }=e_{\alpha }\cdot e_{\beta }$ is the metric tensor
of $\left( d\sigma \right) ^{2}$ \cite{Pa}.

By the well-known \textit{Maupertius-Lagrange's} principle \cite{An,Lu,Mi,Pa}
a path of motion of $\mathcal{\hat{M}}$ in the configurative space is just
the path along which the action is stationary, more precisely along which
the following condition 
\begin{equation}
\vartriangle \mathcal{S}=\sqrt{2m}\Delta \int_{\sigma \left( t_{1}\right)
}^{\sigma \left( t_{2}\right) }\sqrt{%
\kappa%
-\mathcal{P}}d\sigma =0  \label{20}
\end{equation}
is satisfied, that is, by the relation (\ref{18}), the condition 
\begin{equation}
mc\Delta \int_{w\left( t_{1}\right) }^{w\left( t_{2}\right) }dw=0\text{,}
\label{21}
\end{equation}
where $\Delta $ is the variational operator \cite{An,Mi}.

It is well-known from the tensorial analysis \cite{An} that curves of the action
configurative space, for which the condition (\ref{21}) is satisfied, are
geodesics. In addition, the absolute (\textit{Bianchi's}) derivation $%
d_{w}x^{\beta }$ of the unit tangent vector $u=d_{w}\rho $ of those curves,
in the direction of curves, is equal to zero. In other words, the projection
of $du/dw$ onto the tangent hyper-plane of the dual vector basis $\left\{
a^{\beta }\right\} _{\beta =1}^{n}$ of the action metric form, is equal to
zero. Accordingly, the geodesic equations are 
\begin{equation}
\frac{du}{dw}\cdot a^{\gamma }=d_{w}\left( d_{w}x^{\alpha }a_{\alpha
}\right) \cdot a^{\gamma }=d_{ww}^{2}x^{\alpha }a_{\alpha }\cdot a^{\gamma
}+d_{w}x^{\alpha }d_{w}a_{\alpha }\cdot a^{\gamma }=  \label{22}
\end{equation}
\[
=d_{ww}^{2}x^{\gamma }+\hat{\Gamma}_{\alpha \beta }^{\gamma }d_{w}x^{\alpha
}d_{w}x^{\beta }=0\text{,}
\]
where $d_{ww}^{2}$ denotes $d^{2}/\left( dw\right) ^{2}$, the vectors $%
a_{\alpha }=\partial _{\alpha }\rho \left( x^{\delta }\right) $ are
covariant basic vectors forming the covariant vector basis $\left\{
a_{\alpha }\right\} _{\alpha =1}^{n}$, and the mixed system of values $\hat{%
\Gamma}_{\alpha \beta }^{\gamma }=\partial _{\beta }a_{\alpha }\cdot
a^{\gamma }=a^{\gamma \delta }\left( \partial _{\beta }a_{\alpha \delta
}+\partial _{\alpha }a_{\beta \delta }-\partial _{\delta }a_{\alpha \beta
}\right) /2$ are the second order \textit{Christoffel's} symbols \cite{An,Pa}%
.

In the case when $\mathbf{F}$ is a potential force, that is $\mathbf{F}=-grad%
\mathcal{P}$, it follows from the condition $\left( 
\kappa%
-\mathcal{P}\right) e_{\alpha \beta }=ka_{\alpha \beta }$, see (\ref{19})%
\textit{,} that \cite{An} 
\[
\hat{\Gamma}_{\alpha \beta }^{\gamma }=\Gamma _{\alpha \beta }^{\gamma }-%
\frac{1}{2\left( 
\kappa%
-\mathcal{P}\right) }\left( \partial _{\alpha }\mathcal{P}\delta _{\beta
}^{\gamma }+\partial _{\beta }\mathcal{P}\delta _{\alpha }^{\gamma
}-e^{\gamma \delta }e_{\alpha \beta }\partial _{\delta }\mathcal{P}\right) 
\text{,}
\]
since $\Gamma _{\alpha \beta }^{\gamma }=e^{\gamma \delta }\left( \partial
_{\beta }e_{\alpha \delta }+\partial _{\alpha }e_{\beta \delta }-\partial
_{\delta }e_{\alpha \beta }\right) /2$. The new form of the geodesic
equations, for a constrained material point $\mathcal{M}$ ($%
\mathbf{F}\neq \mathbf{0}\Leftrightarrow \mathcal{P}\neq const.$), is 
\begin{equation}
d_{ww}^{2}x^{\gamma }+\Gamma _{\alpha \beta }^{\gamma }d_{w}x^{\alpha
}d_{w}x^{\beta }=\frac{1}{%
\kappa%
-\mathcal{P}}\partial _{\delta }\mathcal{P}d_{w}x^{\delta }d_{w}x^{\gamma }-%
\frac{k}{2\left( 
\kappa%
-\mathcal{P}\right) ^{2}}\partial _{\delta }\mathcal{P}e^{\gamma \delta }%
\text{,}  \label{23}
\end{equation}
that is 
\begin{equation}
m\left( d_{tt}^{2}x^{\gamma }+\Gamma _{\alpha \beta }^{\gamma
}d_{t}x^{\alpha }d_{t}x^{\beta }\right) =-\partial _{\delta }\mathcal{P}%
e^{\gamma \delta }=F^{\gamma }\text{,}  \label{24}
\end{equation}
since $d_{ww}^{2}x^{\gamma }\left( d_{t}w\right) ^{2}+d_{t}x^{\gamma
}d_{w}td_{tt}^{2}w=d_{tt}^{2}x^{\gamma }$, $d_{t}w=c\left( 
\kappa%
-\mathcal{P}\right) /k$ and $d_{tt}^{2}wd_{w}t=-\left[ \partial _{\delta }%
\mathcal{P}/\left( 
\kappa%
-\mathcal{P}\right) \right] d_{t}x^{\delta }$.

So, (\ref{24}) represents \textit{Euler-Lagrange's} differential equations
of extreme curves in the explicit form \cite{Mi}, and at the same time 
\textit{Newton's} low in the field of action of a potential force $%
F^{k}=-\partial _{l}\mathcal{P}e^{kl}$ in the contravariant form \cite{An,Mi}%
: 
\begin{equation}
m(d_{tt}^{2}x^{k}+\Gamma _{ij}^{k}d_{t}x^{i}d_{t}x^{j})=-\partial _{l}%
\mathcal{P}e^{kl}=F^{k}\text{.}  \label{25}
\end{equation}

Accordingly, one may conclude that the dynamical (\textit{Newton's})
equations (\ref{25}) are formally derived from the geometric equations (\ref
{22}) representing \textit{the kinetic energy theorem} on the one hand and 
\textit{Assumption \ref{Pr}} on the other.

In the case of a free material point $\mathcal{M}$, when $\mathcal{P}$ is
constant\textit{:} $\mathcal{P}=const.$, both the basic and action metric
form of the configurative space are \textit{pseudo-euclidean}, while
integral curves are straight-lines \cite{An}. To prove these facts one
starts with \textit{Euler-Lagrange's} differential equations 
\begin{equation}
d_{w}\left( \partial _{d_{w}x^{\beta }}\mathcal{W}\right) -\partial _{\beta }%
\mathcal{W}=0\text{,}  \label{26}
\end{equation}
where 
\begin{equation}
\mathcal{W}=\frac{%
\kappa%
-\mathcal{P}}{k}e_{\alpha \beta }d_{w}x^{\alpha }d_{w}x^{\beta }\text{,}
\label{27}
\end{equation}
as the condition for the action (\ref{21}) to be stationary. The geodesic
equations (\ref{22}) are explicitly obtained from it in a known way. If
spatial co-ordinates of the configurative space are spherical ones ($%
r,\theta ,\varphi $), then the components of $e_{\alpha \beta }$ are only
functions of the co-ordinates $r$ and $\varphi $ \cite{N-V}, so that it
follows from (\ref{26}) that 
\begin{equation}
d_{w}(\frac{%
\kappa%
-\mathcal{P}}{k}e_{11}cd_{w}t)=0  \label{28}
\end{equation}
and 
\begin{equation}
d_{w}(\frac{%
\kappa%
-\mathcal{P}}{k}e_{33}d_{w}\theta )=0\text{,}  \label{29}
\end{equation}
that is 
\begin{equation}
\left( 
\kappa%
-\mathcal{P}\right) cd_{w}t=k  \label{30}
\end{equation}
and 
\begin{equation}
\left( 
\kappa%
-\mathcal{P}\right) \left( r^{2}\cos ^{2}\varphi \right) d_{w}\theta
=k\alpha \text{.}  \label{31}
\end{equation}

Let the polar extension $r\,$and the polar angle $\theta $ be intensities of 
$\mathbf{r}$ and an angle between the position vector $\mathbf{r}$ and the
polar axis $p$ passing through the origin and the perihelial point,
respectively. Then, since $S=r^{2}\dot{\theta}=const.$, where $\mathbf{S}=%
\mathbf{r}\times \mathbf{v}$ is the sector velocity vector \cite{Ra}, it
follows from the condition (\ref{31}) that a free material point motion is
the plane motion ($\varphi =0$) and $S=\alpha c$. As $\left( ds\right)
^{2}=\left( dr\right) ^{2}+r^{2}\left( d\theta \right) ^{2}$ then we obtain
finally from (\ref{9}), (\ref{19}) and (\ref{31}) that 
\begin{equation}
\left( dr\right) ^{2}=r^{4}[(\frac{1}{\alpha })^{2}-(\frac{1}{r})^{2}+\frac{%
\kappa%
-\mathcal{P}}{k\alpha ^{2}}]\left( d\theta \right) ^{2}\text{,}  \label{32}
\end{equation}
that is 
\begin{equation}
d_{\theta \theta }^{2}\frac{1}{r}+\frac{1}{r}=0\text{,}  \label{33}
\end{equation}
and what is just \textit{Binet's} differential equation \cite{Mi}. The
solution to this differential equation defines a straight-line in the polar
co-ordinates\textit{:} $r_{0}=r\cos \theta $, where $r_{0}$ is the
perihelial distance.

If (\ref{15}) is analyzed anew, in the case when the upper limit of
integration is changeable ($t_{2}=t$) 
\begin{equation}
\mathcal{S}=2\int_{t_{1}}^{t}\mathcal{K}dt\text{,}  \label{34}
\end{equation}
then considering the fact that 
\begin{equation}
d_{t}\mathcal{S}=2\mathcal{K}=me_{\alpha \beta }d_{t}x^{\alpha
}d_{t}x^{\beta }\text{,}  \label{35}
\end{equation}
it is easy to see that from great interest in the analysis is the functional 
$\mathcal{S}^{H}$ \cite{Mi}, nominally equal to $\mathcal{S}$, for which the
following functional relations hold: 
\begin{equation}
\partial _{\alpha }\mathcal{S}^{H}=me_{\alpha \beta }d_{t}x^{\beta }
\label{36}
\end{equation}
and 
\begin{equation}
d_{t}\mathcal{S}^{H}=2\mathcal{K}\text{,}  \label{37}
\end{equation}
as well as the functional $\mathcal{Z}^{H}$ satisfying the condition 
\begin{equation}
\mathcal{S}^{H}=\mathcal{Z}^{H}-(\frac{1}{2}mc^{2}-%
\kappa%
)t\text{,}  \label{38}
\end{equation}
that is, the condition 
\begin{equation}
d_{t}\mathcal{Z}^{H}=\mathcal{L}\text{,}  \label{39}
\end{equation}
since $d_{t}\mathcal{S}^{H}=2\left( 
\kappa%
-\mathcal{P}\right) $.

The functional $\mathcal{L}=\mathcal{E}-\mathcal{P}$ is \textit{Lagrange's}
functional and in the acute case of the standard \textit{Lagrange's} system
it is also \textit{Lagrangian} of $\mathcal{M}$ \cite{Mi}.

Since $\partial _{t}\mathcal{S}^{H}=-mc^{2}$ for $x^{4}=ct$, see (\ref{36}),
it follows from the condition (\ref{38}) that $\partial _{t}\mathcal{Z}^{H}=-%
\mathcal{U}$, that is 
\begin{equation}
\partial _{t}\mathcal{Z}^{H}+\mathcal{H}=0\text{.}  \label{40}
\end{equation}

The preceding equation is \textit{Hamilton-Jacobi's} equation and according
to that the functional $\mathcal{Z}^{H}$ is the principal \textit{Hamilton's}
functional of $\mathcal{M}$ \cite{Mi}. The \textit{Hamiltonian} (\textit{%
Hamilton's} functional) $\mathcal{H}$ of $\mathcal{M}$ is equal to the
functional of the total mechanical energy $\mathcal{U}$, more precisely to
an integral of motion, and what in accordance with the fact that the kinetic
energy $\mathcal{E}$ of $\mathcal{M}$ is a homogenous square function of $%
d_{t}x^{\alpha }$ \cite{Mi}.

As $\partial _{i}\mathcal{Z}^{H}=me_{ij}d_{t}x^{j}$, see (\ref{36}) and (\ref
{38}), that is 
\begin{equation}
e^{kl}d_{k}\mathcal{Z}^{H}d_{l}\mathcal{Z}%
^{H}=m^{2}e_{ij}d_{t}x^{i}d_{t}x^{j}=2m\mathcal{E}\text{,}  \label{41}
\end{equation}
then it follows that 
\begin{equation}
-\partial _{t}\mathcal{Z}^{H}-\frac{1}{2m}e^{kl}d_{k}\mathcal{Z}^{H}d_{l}%
\mathcal{Z}^{H}=\mathcal{P}\text{,}  \label{42}
\end{equation}
what is only the second form of \textit{Hamilton-Jacobi's} equation.

\subsection{Principle of invariance and co-ordinate transformations}

In \textit{Section 1} the mathematical model of a material point motion in
the configurative space (in the spatial subspace of the space-time
continuum) and in the field of action of an active potential force $\mathbf{F%
}$ with the potential $\mathcal{P}$ has been derived on the basis of \textit{%
the kinetic energy theorem} and \textit{Assumption \ref{Pr}}. Note anew that
the basic metric form $\left( d\sigma \right) ^{2}$ represents \textit{the
kinetic energy theorem.} In addition, $\left( ds\right) ^{2}$ is explicitly
related to the kinetic energy of $\mathcal{M}$.

\begin{description}
\item[Principle of invariance]  \label{Prin}\textit{Any form of energy of }$%
\mathcal{M}$\textit{\ is an invariant in the more expansive sense (a scalar
invariant), more precisely a zero order tensor}.$\blacktriangledown $
\end{description}

Co-ordinate systems of configurative spaces are only auxiliary tools for
analysis, so that, on the one hand, all fundamental values characterizing
inner (natural) dynamical properties of a system, as in this case ether the
potential or kinetic energy of $\mathcal{M}$, are invariants with respect to
co-ordinate transformations \cite{An}. Accordingly, the fundamental
differential functional form of \textit{the kinetic energy theorem} is also
an absolute invariant (an invariant functional form), so that on the basis
of these two principles of invariance we can say that the time is also
tensor invariant. In other words, in any system of co-ordinate
transformations of configurative spaces the time is absolutely only one
independent variable.

At the choice of co-ordinate transformations of ether basic or action metric
forms of configurative spaces it is necessary in addition to the condition
for the so called \textit{Jacobian} of transformation to be different from
zero to take also \textit{the principle of invariance \ref{Prin}} into
consideration. It is different from co-ordinate transformations of ether
basic or action metric forms of configurative spaces at the choice of which
the condition for \textit{Jacobian} of transformation to be different from
zero has to be satisfied only. Hence, it is clear that the co-ordinate
transformations, as \textit{Lorentz's} transformations \cite{Gr,Iv,L-L,Pa},
do not meet the established criteria for the choice of co-ordinate
transformations of ether basic or action metric forms of configurative
spaces, and for reason that \textit{the principle of invariance \ref{Prin} }%
is being destroyed by them.

\subsection{Example\textit{: The mathematical model of two material points
motion in the field of action of the central conservative force, as an
idealization of two body problem}}

Let $\mathbf{v}_{1}=d_{t}\mathbf{r}_{1}$ and $\mathbf{v}_{2}=d_{t}\mathbf{r}%
_{2}$ be velocity vectors of material points $\mathcal{M}_{1}$ and $\mathcal{%
M}_{2}$, respectively, in a spatial subspace of the space-time continuum.
Then, a motion analysis of $\mathcal{M}_{1}$ and $\mathcal{M}_{1}$ reduced
to that of the compound motion of a virtual material point of the mass $\mu
=\left( m_{1}m_{2}\right) /M$ ($M=m_{1}+m_{2}$) in the configurative space
of the spatial continuum. This is based on defined property of the central
conservative force $\mathbf{F}$ that its direction of assaulted action
coincides with the direction of the relative position vector $\mathbf{r}=%
\mathbf{r}_{1}-\mathbf{r}_{2}$ of $\mathcal{M}_{1}$ and $\mathcal{M}_{2}$,
as well as on \textit{Newton's} low of motion and the moment low of quantity
of movement \cite{Mi,Ra}. Having in view the fact that an absolute motion of
the mass centre of $\mathcal{M}_{1}$ and $\mathcal{M}_{2}$ is uniformly \cite
{Mi} the mathematical model of motion of $\mathcal{M}_{1}$ and $\mathcal{M}%
_{2}$ in the configurative space of the spatial continuum and in the field
of action of the central conservative force $\mathbf{F}$ can be derived as
follows. This mathematical model cams from that of motion of a virtual
material point $\mathcal{M}_{\mu }$ in an immovable configurative space of
the spatial continuum and in the field of action of the central conservative
force $\mathbf{F}$ whose assaulted direction of action coincides with the
direction of the position vector $\mathbf{r}$ of $\mathcal{M}_{\mu }$ with
respect to the immovable centre of mass as the origin. Clearly, in spite of
all that the relativistic principle of the classical mechanics that says
that a relative movement in an inertial co-ordinate system is analogous to
an absolute movement in an immovable co-ordinate system \cite{Ra} will be a
base for further analysis.

As, in this case too, the sector velocity vector $\mathbf{S}=\mathbf{r}%
\times \mathbf{v}$ is obviously constant, then it follows from (\ref{19}): 
\begin{equation}
k\left( dw\right) ^{2}=\left( 
\kappa%
-\mathcal{P}\right) \left( d\sigma \right) ^{2}=\left( 
\kappa%
-\mathcal{P}\right) [\left( dr\right) ^{2}+r^{2}\left( d\theta \right)
^{2}-c^{2}\left( dt\right) ^{2}]\text{,}  \label{43}
\end{equation}
as well as from the geodesic equations, see (\ref{28}) and (\ref{29}): 
\begin{equation}
d_{w}(\frac{%
\kappa%
-\mathcal{P}}{k}cd_{w}t)=0  \label{44}
\end{equation}
and 
\begin{equation}
d_{w}(\frac{%
\kappa%
-\mathcal{P}}{k}r^{2}d_{w}\theta )=0\text{,}  \label{45}
\end{equation}
more precisely from (\ref{32}): 
\begin{equation}
\left( dr\right) ^{2}=r^{4}[(\frac{1}{\alpha })^{2}-(\frac{1}{r})^{2}+\frac{%
\kappa%
-\mathcal{P}}{k\alpha ^{2}}]\left( d\theta \right) ^{2}\text{,}  \label{46}
\end{equation}
that 
\begin{equation}
d_{\theta \theta }^{2}\frac{1}{r}+\frac{1}{r}=-\frac{1}{2k\alpha ^{2}}%
\partial _{\frac{1}{r}}\mathcal{P}\text{,}  \label{47}
\end{equation}
what is the well-known \textit{Binet's} differential equation of the
trajectory of motion of $\mathcal{M}_{\mu }$ in an immovable configurative
space of the spatial continuum.

\subsection{Modified \textit{Newton's} gravity concept}

For the conservative \textit{Newton's} gravity force $\mathbf{F}_{N}=-grad%
\mathcal{P}$ the expression $%
\kappa%
-\mathcal{P}$ is of the following functional form $%
\kappa%
-\mathcal{P}=-k(\hat{\lambda}-2\varrho /r)$, where $k=\mu c^{2}/2$, $\hat{%
\lambda}$ is an integral constant of initial conditions, and $\varrho $ is a
constant of the gravitational radius\textit{:} $\varrho =\gamma M/c^{2}$ ($%
\gamma $ is the well-known gravitational constant), so that (\ref{47}) is
reduced to \textit{Binet's} differential equation of motion of a constrained
material point $\mathcal{M}_{\mu }$ in an immovable configurative space of
the spatial continuum and in the field of action of the central \textit{%
Newton's} gravity force $\mathbf{F}_{N}$: 
\begin{equation}
d_{\theta \theta }^{2}\frac{1}{r}+\frac{1}{r}=\frac{\varrho }{\alpha ^{2}}%
\text{,}  \label{48}
\end{equation}
where $\alpha =S/c$ and $S=r^{2}d_{t}\theta =const$.

Since in the limit as $r\rightarrow 0^{+}$ the coefficient of the
proportionality between the action $\left( dw\right) ^{2}$ and basic $\left(
d\sigma \right) ^{2}$ metric forms: 
\begin{equation}
\hat{\lambda}-\frac{2\varrho }{r}=\lambda +1-\frac{2\varrho }{r}\text{,}
\label{49}
\end{equation}
where $\lambda =\hat{\lambda}-1$, tends to infinity, it is logical to assume
that the last two terms on the right hand side of (\ref{49}) approximately
represent an expansion of the exponential function $e^{-2\varrho /r}$ into 
\textit{Taylor's} functional series. In other words, it is logical to assume
that the real functional coefficient of the proportionality between the
action $\left( dw\right) ^{2}$ and basic $\left( d\sigma \right) ^{2}$
metric forms, in the case when $\mathcal{M}_{\mu }$ moves in the field of
action of the modified central \textit{Newton's} gravity force $\mathbf{%
\digamma }_{N}$, is as follows 
\begin{equation}
\left( dw\right) ^{2}=-(\lambda +e^{-\frac{2\varrho }{r}})\left( d\sigma
\right) ^{2}\text{.}  \label{50}
\end{equation}

Accordingly, the modified \textit{Binet's} differential equation of motion
of $\mathcal{M}_{\mu }$ in the field of action of the modified central
conservative \textit{Newton's} gravity force $\mathbf{\digamma }_{N}$: 
\begin{equation}
\mathbf{\digamma }_{N}\,\mathbf{=}-grad\mathcal{P}\left( \mathbf{r}\right) =-%
\frac{2k\varrho }{r^{3}}e^{-\frac{2\varrho }{r}}\mathbf{r}\text{,}
\label{51}
\end{equation}
with the potential $\mathcal{P}\left( \mathbf{r}\right) $\textit{:} $%
\kappa%
-\mathcal{P}=-k(\lambda +e^{-2\varrho /r})$, is the following differential
form 
\begin{equation}
d_{\theta \theta }^{2}\frac{1}{r}+\frac{1}{r}=\frac{\varrho }{\alpha ^{2}}%
e^{-\frac{2\varrho }{r}}\text{.}  \label{52}
\end{equation}

\subsubsection{A material point motion in the field of action of$\ \mathbf{%
\digamma }_{N}$}

Start with the differential equation of the second \textit{Newton's} law
(principle) of a material point motion \cite{Mi,N-V} 
\begin{equation}
\mu d_{tt}^{2}\mathbf{r}=-\frac{2k\varrho }{r^{3}}e^{-\frac{2\varrho }{r}}%
\mathbf{r}\text{.}  \label{53}
\end{equation}

Multiplying the preceding differential equation on the right by the sector
velocity vector $\mathbf{S}=\mathbf{r\times v}$: 
\begin{equation}
\mu d_{tt}^{2}\mathbf{r\times S=}-\frac{2k\varrho }{r^{3}}e^{-\frac{2\varrho 
}{r}}\mathbf{r\times S}\text{,}  \label{54}
\end{equation}
we obtain that 
\begin{equation}
d_{t}\left( \mathbf{v\times S}\right) =\gamma Me^{-\frac{2\varrho }{r}}d_{t}%
\frac{\mathbf{r}}{r}\text{,}  \label{55}
\end{equation}
since $d_{t}\mathbf{S}=\mathbf{0}$, that is 
\begin{equation}
d(\mathbf{v\times S}-\gamma M\mathbf{r}_{0})=\gamma M(e^{-\frac{2\varrho }{r}%
}-1)d\mathbf{r}_{0}\text{.}  \label{56}
\end{equation}

By (\ref{46}): 
\begin{equation}
\left( d_{\theta }r\right) ^{2}=r^{4}[(\frac{1}{\alpha })^{2}-(\frac{1}{r}%
)^{2}-\frac{1}{\alpha ^{2}}(\lambda +e^{-\frac{2\varrho }{r}})]\text{,}
\label{57}
\end{equation}
it follows obviously that 
\begin{equation}
\lambda +e^{-\frac{2\varrho }{r}}=1-\alpha ^{2}[(d_{\theta }\frac{1}{r}%
)^{2}+(\frac{1}{r})^{2}]=1-\frac{v^{2}}{c^{2}}\text{.}  \label{58}
\end{equation}

Accordingly, based on (\ref{56}), there holds 
\begin{equation}
d\mathbf{L=}\gamma M(e^{-\frac{2\varrho }{r}}-1)d\mathbf{r}_{0}=-\gamma
M(\lambda +\frac{v^{2}}{c^{2}})d\mathbf{r}_{0}\text{,}  \label{59}
\end{equation}
where the vector 
\begin{equation}
\mathbf{L=v\times S}-\gamma M\mathbf{r}_{0}\text{,}  \label{60}
\end{equation}
satisfying the relations: $\mathbf{L\cdot S}=0$ and $\mathbf{L\cdot L}%
=L^{2}=v^{2}S^{2}-2\gamma M/r+\gamma ^{2}M^{2}$, in this acute case is not
more an element of \textit{Milankovi\'{c}'s} constant vector elements, more
precisely is not more \textit{Laplace's} integration vector constant \cite
{Mih}.

By multiplying (\ref{60}) by $\mathbf{r}$: $\mathbf{L\cdot r=}\left( \mathbf{%
v\times S}\right) \mathbf{\cdot r}-\gamma M\mathbf{r}$, and considering the
fact that $\left( \mathbf{v\times S}\right) \mathbf{\cdot r}=S^{2}$, the
equation of the trajectory of motion of $\mathcal{M}_{\mu }$ in the field of
action of the modified \textit{Newton's} gravity force $\mathbf{\digamma }%
_{N}$ is reduced to 
\begin{equation}
r=\frac{1}{\gamma M}\frac{S^{2}}{1+\frac{L}{\gamma M}\cos \varphi }\text{,}
\label{61}
\end{equation}
where $\varphi $ is an angle between the vectors $\mathbf{r}$ and $\mathbf{L}
$.

On the basis of the preceding equalities the following functional equalities
are easily obtained 
\begin{equation}
\frac{v^{2}}{c^{2}}=\frac{\varrho ^{2}}{\alpha ^{2}}[1+(\frac{L}{\gamma M}%
)^{2}+\frac{2L}{\gamma M}\cos \varphi ]  \label{62}
\end{equation}
and 
\begin{equation}
\frac{2\varrho }{r}=\frac{2\varrho ^{2}}{\alpha ^{2}}(1+\frac{L}{\gamma M}%
\cos \varphi )\text{,}  \label{63}
\end{equation}
as well as 
\begin{equation}
d\mathbf{L}=-\gamma M\{\lambda +\frac{\varrho ^{2}}{\alpha ^{2}}[1+(\frac{L}{%
\gamma M})^{2}+\frac{2L}{\gamma M}\cos \varphi \}]d\mathbf{r}_{0}  \label{64}
\end{equation}
and 
\begin{equation}
d\mathbf{L}=\gamma M[e^{-\frac{2\varrho ^{2}}{\alpha ^{2}}(1+\frac{L}{\gamma
M}\cos \varphi )}-1]d\mathbf{r}_{0}\text{.}  \label{65}
\end{equation}

As $\theta =\varphi +\omega $ and $d\mathbf{r}_{0}=d\theta \mathbf{t}$,
where $\omega $ is an angle of deviation of $\mathbf{L}$ with respect to the
perihelial direction: $d\left( \mathbf{L}/L\right) =d\mathbf{L}_{0}=d\omega 
\mathbf{k}$, when $\varphi =0$, while $\mathbf{t}$ and $\mathbf{k}$ are unit
vectors being orthogonal onto the unit vectors $\mathbf{r}_{0}$ and $\mathbf{%
L}_{0}$, respectively, multiplying (\ref{64}) and (\ref{65}) by $\mathbf{k}$%
, we obtain finally that 
\begin{equation}
d\omega =-\frac{\gamma M}{L}\{\lambda +\frac{\varrho ^{2}}{\alpha ^{2}}[1+(%
\frac{L}{\gamma M})^{2}+\frac{2L}{\gamma M}\cos \varphi ]\}\cos \varphi
\left( d\varphi +d\omega \right)  \label{66}
\end{equation}
and 
\begin{equation}
d\omega =\frac{\gamma M}{L}[e^{-\frac{2\varrho ^{2}}{\alpha ^{2}}(1+\frac{L}{%
\gamma M}\cos \varphi )}-1]\cos \varphi \left( d\varphi +d\omega \right) 
\text{.}  \label{67}
\end{equation}

Accordingly, approximative values of the ratio of angular speeds of $\mathbf{%
L}$ and $\mathbf{r}$, at the perihelion ($\varphi =0$) and aphelian ($%
\varphi =\pi $), are 
\begin{equation}
\frac{d\omega }{d\varphi }\left| _{\varphi =0}\right. \approx -\frac{%
2\varrho ^{2}}{\alpha ^{2}}(1+\frac{\gamma M}{L})\approx -\frac{2\varrho 
\frac{\gamma M}{L}}{a(1-\frac{L}{\gamma M})}  \label{68}
\end{equation}
and 
\begin{equation}
\frac{d\omega }{d\varphi }\left| _{\varphi =\pi }\right. \approx -\frac{%
2\varrho ^{2}}{\alpha ^{2}}(1-\frac{\gamma M}{L})\approx \frac{2\varrho 
\frac{\gamma M}{L}}{a(1+\frac{L}{\gamma M})}\text{,}  \label{69}
\end{equation}
where a constant $2a$ is the major axis of an elliptical orbit of $\mathcal{M%
}_{\mu }$ in the field of action of $\mathbf{\digamma }_{N}$ \cite{Ra}.

By the middle approximative values of the recessional and precessional
angular deviations of the perihelial point vector per revolution ($\varphi
=2\pi $)\textit{:} 
\begin{equation}
\left( \bar{\omega}\right) _{r}\approx -\frac{\pi \varrho ^{2}}{\alpha ^{2}}%
(1+\frac{\gamma M}{L})\approx -\frac{\pi \varrho \frac{\gamma M}{L}}{a(1-%
\frac{L}{\gamma M})}  \label{70}
\end{equation}
and 
\begin{equation}
\left( \bar{\omega}\right) _{p}\approx -\frac{\pi \varrho ^{2}}{\alpha ^{2}}%
(1-\frac{\gamma M}{L})\approx \frac{\pi \varrho \frac{\gamma M}{L}}{a(1+%
\frac{L}{\gamma M})}\text{,}  \label{71}
\end{equation}
the total middle approximative value of the perihelion advance per
revolution ($\varphi =2\pi $), is 
\begin{equation}
\bar{\omega}=\left( \bar{\omega}\right) _{r}+\left( \bar{\omega}\right)
_{p}\approx -\frac{2\pi \varrho ^{2}}{\alpha ^{2}}\approx -\frac{2\pi
\varrho }{a[1-(\frac{L}{\gamma M})^{2}]}\text{.}  \label{72}
\end{equation}

Note that the exact value of an angular deviation of the perihelial point
vector with respect to the rotation of the position vector $\mathbf{r}$ can
be obtained by integrating (\ref{66}) and (\ref{67}), and what is very
complex problem. Anyway, it is very important that the angular deviation of
the perihelial point vector has been just obtained as a result of the
analysis of a material point motion in the field of action of the
approximately modified \textit{Newton's} gravity force $\mathbf{\digamma }%
_{N}$.

In addition, it is very indicative that curvatures of the \textit{Riemannian}
spaces\textit{:} 
\begin{equation}
dw^{2}=e^{\nu \left( r\right) }dr^{2}+r^{2}d\theta ^{2}+\sin ^{2}\theta
d\varphi ^{2}+e^{-\nu \left( r\right) }dt^{2}\text{,}  \label{73}
\end{equation}
and 
\begin{equation}
d\hat{w}^{2}=e^{\hat{\nu}\left( r\right) }\left( dr^{2}+r^{2}d\theta
^{2}+\sin ^{2}\theta d\varphi ^{2}+dt^{2}\right) \text{\textit{,}}
\label{74}
\end{equation}
where $\nu \left( r\right) =-\ln (1-2\varrho /r)$ and $\hat{\nu}\left(
r\right) =-2\varrho /r$, are approximately equivalent $\kappa \simeq \hat{%
\kappa}$, since (see \cite{Sa 2}) 
\begin{equation}
\kappa =2\frac{\varrho ^{2}}{r^{6}}  \label{75}
\end{equation}
and 
\begin{equation}
\hat{\kappa}=2\frac{\varrho ^{2}}{r^{6}}e^{\frac{4\varrho }{r}}\sqrt{(1+%
\frac{\varrho }{2r})(1+\frac{\varrho }{r})}\text{.}  \label{76}
\end{equation}

Clearly, the first space is the strict \textit{Schwarzshild-Droste's}
solution to the static gravitational field with spherical symmetry, and the
second one is its modification based on the approximately modified \textit{%
Newton's} gravity force $\mathbf{\digamma }_{N}$ (see Figure 1.).

\FRAME{fhFU}{2.4146in}{3.4541in}{0pt}{\Qcb{The modified \textit{Newton's}
gravity force}}{}{fig1.gif}{\special{language "Scientific Word";type
"GRAPHIC";maintain-aspect-ratio TRUE;display "USEDEF";valid_file "F";width
2.4146in;height 3.4541in;depth 0pt;original-width 7.8923in;original-height
11.3334in;cropleft "0";croptop "1";cropright "1";cropbottom "0";filename
'fig1.gif';file-properties "XNPEU";}}

\section{Conclusion}

The mathematical model of a material point motion in the three-dimensional
spatial subspace of the four-dimensional integral space-time continuum and
in the field of action of a conservative active force $\mathbf{F}$ is
analogous to \textit{Newton's} mathematical model of the classical
mechanics. In addition, the configurative integral space of the space-time
continuum, whose the metric form $\left( d\sigma \right) ^{2}$ represents
one of the fundamental theorems of the material point dynamics \textit{the
kinetic energy theorem} as well as \textit{Assumption \ref{Pr}}, is the
configurative space of the space-time continuum of \textit{Minkowski} from 
\textit{Einstein's} relativity theory. Accordingly, it can be said that in
the paper a new connection has been established, in contrast to an
approximative one, between the classical \textit{Newton's} mathematical
model and the relativistic \textit{Einstein's} mathematical model.

On the other hand the approximately modified \textit{Newton's} gravity
concept is not, from any point of view, in collision with old \textit{%
Newton's} one. At the same time it solves the acutely vexed questions within
old \textit{Newton's} gravity concept (the singularity and perihelion
problems). Furthermore, analyzing the analytical expression for the modified 
\textit{Newton's} gravity force $\mathbf{\digamma }_{N}$, see (\ref{51}), we
can separate the four indicative domains of its field of action (Fig. 1.).
The first one is a domain of the weak action on finitely small distances.
The second one is a domain of the strong action in a neighborhood of the
gravitational radius $\varrho =\gamma M/c^{2}$ ($\partial _{r}F\left|
_{r=\varrho }\right. =0$ and $\partial _{rr}^{2}F\left| _{r=\varrho }\right.
<0$). The third one is a domain of action on finitely large distances
relative to the gravitational radius $\varrho $ and with the relatively
small velocities relative to the light velocity, and the fourth on finitely
large distances relative to the gravitational radius $\varrho $ and with
velocities that are comparable to the light velocity. Previously separated
domains of the field of action of the modified \textit{Newton's} gravity
force $\mathbf{\digamma }_{N}$ it would be desirable to compare to the
fields of action of the four so far non-unified fundamental forces (weak and
strong nuclear interactions, gravity and \textit{Lorenz's}
electromagnetism), and what could be the subject of a separate analysis.
Accordingly, note at the end that a correction to \textit{Newton's} gravity
law in the form of the functional dependence $r^{-3}e^{-r/r_{c}}$ $\mathbf{r}
$ irresistibly reminding of the modified \textit{Newton's} gravity force,
see (\ref{51}), and obviously wrongly called the ''fifth force'', has been
revealed by a reexamination of the old attraction data and careful new force
measurements \cite{BBFT}.

\end{document}